\documentclass[12pt]{article}

\usepackage{amsmath}
\usepackage{epsfig}
\usepackage{amstext,amssymb,amsfonts}

\advance\voffset by -2.0cm
\advance\hoffset by -1.75cm
\def\Tr{\,{\rm Tr}\, }
\def\Trqy{\,{\rm Tr}_\R^{q,y}\, }

\def\be{\begin{equation}}
\def\ee{\end{equation}}
\def\ba{\begin{eqnarray}}
\def\ea{\end{eqnarray}}

\newcommand{\G}{{\cal G}}
\renewcommand{\H}{{\cal H}}

\newcommand{\N}{{\cal N}}
\newcommand{\R}{{\cal R}}

\newcommand{\ct}{\tilde{c}}

\newcommand{\ie}{{\it i.e.~}}

\newcommand{\Zop}{\mathbb{Z}}
\newcommand{\Nop}{\mathbb{N}}

\title{\LARGE Differential operators for elliptic genera}

\author{
Matthias R.\ Gaberdiel$^{1}$\thanks{\tt E-mail: gaberdiel@itp.phys.ethz.ch}\hspace{0.2cm}
and
Christoph A.\ Keller$^{2}$\thanks{\tt E-mail: ckeller@physics.harvard.edu} \\
\\ \\
${}^{1}${\small Institut f\"ur Theoretische Physik, 
ETH Z\"urich} \vspace*{-0.1cm} \\
{\small 8093 Z\"urich, Switzerland} \vspace{0.3cm} \\
${}^{2}${\small Jefferson Physical Laboratory, Harvard University}
\vspace*{-0.1cm} \\
{\small Cambridge, MA 02138, USA}
}
\date{\today}
\begin{document}
\maketitle

\begin{abstract}
Using the generalisation of Zhu's recursion relations to $N=2$ superconformal field
theories we construct modular covariant differential operators for weak
Jacobi forms. We show that differential operators of this type characterise
the elliptic genera of $N=2$ superconformal minimal models, and 
sketch how they can be used to constrain extremal 
$N=2$ superconformal field theories.
\end{abstract}

\renewcommand{\theequation}{\arabic{section}.\arabic{equation}}


\section{Introduction}
It is well known that there is a deep connection between
(rational) conformal field theory and the modular group. The origin of this relation
lies in the fact that the correlation functions of a conformal field theory 
on a torus can only depend on its conformal structure, which
is parametrised by elements in the quotient space 
$\mathbb{H}^+/{\rm SL}(2,\mathbb{Z})$. Formulated as a function of the modular
parameter $\tau\in \mathbb{H}^+$, the correlation functions must therefore be
modular covariant, {\it i.e.}\ covariant under the action of ${\rm SL}(2,\mathbb{Z})$.  

The full torus amplitude can be expressed in terms of 
the chiral characters of the conformal field theory; from this point of view
the modular covariance of the correlation functions then comes from the property
of these characters to form a vector valued modular form. The corresponding
representation of the modular group, in particular the modular $S$-matrix, encodes
important information about the structure of the conformal field theory,
for example it determines the fusion rules via the Verlinde formula 
\cite{Verlinde:1988sn}.

The mathematical argument establishing the modular covariance properties 
of the characters (under some weak assumptions) was given some time ago by Zhu
\cite{Zhu}. A key step in his argument involved the construction of a modular
differential equation that annihilates all the characters of a given conformal field theory. 
(Modular differential equations were first considered from a slightly different point
of view in \cite{MMS,MMS1}, see also \cite{Eguchi:1986sb,Anderson:1987ge}.)
This differential equation always comes from a null vector in the 
vacuum representation \cite{Gaberdiel:2007ve,Gaberdiel:2008pr}. 
One can also turn the logic around and construct  differential operators
starting from essentially arbitrary vectors in the vacuum representation. 
The resulting
differential operators are automatically modular covariant; in fact, they are simply 
the well known holomorphic modular differential operators that can be obtained 
by suitable iterations of the operator
\be
D_s = q\frac{d}{dq} - \frac{s}{4\pi^2}G_2(q)\ ,
\ee
where $G_2(q)$ is the second Eisenstein series.
\smallskip

For $N=2$ superconformal field theories the natural generalisation of the chiral 
characters are the elliptic genera. These also have good modular properties; in
fact, they define vector
valued weak Jacobi forms \cite{Kawai:1993jk}. (From a 
mathematical point of view this was recently established in \cite{Eicher}.) For weak 
Jacobi forms, however, much less is known about modular differential 
operators. The purpose of this note is to construct 
such operators by generalising the above method to the $N=2$ case.
As we shall see, one is  naturally led to introduce `twisted' Eisenstein series
\cite{Mason:2008zz} (see also \cite{Dong:1997ea}) in the process.
These twisted Eisenstein series serve the same purpose
as $G_2$ in the bosonic example above, \ie they cancel
the anomalies introduced by the derivatives with
respect to $q$ and $y$. Since their transformation properties
under the modular group are much more complicated, it
is far from obvious how to combine them with derivatives
to obtain modular covariant operators. The approach
presented in this note gives a procedure which automatically
leads to modular covariant combinations of (twisted) Eisenstein 
series and derivatives.
\medskip

Physically the interest in analysing modular differential
operators comes from the proposed existence of extremal
conformal field theories. A conformal field theory cannot 
just contain the vacuum and its (super-)Virasoro descendants since 
the resulting partition function would not be modular invariant. It is
thus necessary to have additional primary fields, and modularity
gives an upper bound for the weight of the lowest of these
new fields. The analysis for the holomorphic \cite{Hoehn,Witten:2007kt}, 
non-holomorphic \cite{Hellerman:2009bu}, 
and $N=2$ supersymmetric \cite{Gaberdiel:2008xb} cases always 
gives a bound linear in $c$, where the coefficient depends on the 
additional assumptions. Nonetheless, finding a partition function is only
a necessary condition for the existence of a conformal field theory, and
it is not clear whether these modular functions are in fact 
partition functions of consistent conformal field theories.
One is thus motivated
to look for additional conditions that have to be
satisfied. In \cite{Gaberdiel:2007ve,Gaberdiel:2008pr}
such a constraint was derived: given the characters of
a putative conformal field theory, it is always possible to construct by
inspection a modular differential equation which annihilates all of 
them. From the existence of this differential equation
one can then deduce a null vector relation in the vacuum representation,
{\it i.e.}\ obtain insights into some other part of the structure of the theory.
A necessary prerequisite for this approach is to have some control
over the structure of the modular differential operators. The results
of this note should therefore allow one to make progress in this
direction for $N=2$ theories; this is briefly sketched in section~4.2.
\medskip

The remainder of the note is organised as follows. In the following
section we explain the structure of the recursion relations for the
elliptic genus amplitudes. In section~3 we then find a family of 
differential operators that map weak Jacobi forms of weight zero and arbitrary index 
to weak Jacobi forms of higher weight and the same index. In section~4 we apply
these differential operators to the elliptic genera of $N=2$ 
minimal models and to extremal $N=2$ conformal
field theories. There are a number of appendices where some
of the more technical calculations are described.

\section{Recursion relations for elliptic genera}
\setcounter{equation}{0}

In the following we want to derive differential equations
for the elliptic genus of an $N=2$ superconformal field theory.
The differential equations can be obtained  using analogous arguments as those
used by Zhu in the analysis of the characters \cite{Zhu}. Below we
shall summarise the relevant formulae; the derivations and a summary of our conventions 
can be found in appendix~\ref{app:zhuN2}.
\smallskip

Let us consider the functions
\be
\Trqy({\cal O} )= \Tr_\R ( {\cal O} \, q^{L_0-\frac{c}{24}} \, y^{J_0}\, (-1)^F)\ ,
\ee 
where $\R$ is the (Ramond) representation over which we take the trace, 
$J_0$ is the zero mode of the $U(1)$ current in the $N=2$ superconformal
algebra, $F$ is the fermion number operator, and ${\cal O}$ is an operator
from the vertex operator algebra that we insert. 
The main idea of the analysis is to derive recursion relations for the
functions $\Trqy({\cal O})$. For example, if $a$ is an element of the vertex 
operator algebra whose $U(1)$ charge is zero, $J_0 a = 0$, then
we have the original formulae of Zhu for the bosonic case,
\be
{}\Trqy \bigl( o(a_{[-h_a]}b)\bigr) = 
\Trqy \bigl(o(a_{[-h_a]} \Omega) \,o(b)\bigr)
+ \sum_{k=1}^\infty G_{2k}(q) 
    \Trqy \bigl( o(a_{[2k-h_a]}b) \bigr)\ , \label{rec01}
\ee
and
\be \label{rec02}
{}\Trqy \bigl(o(a_{[-h_a-n]}b) \bigr)= (-1)^n \sum_{2k\geq n+1} \binom{2k-1}{n}
G_{2k}(q)\, \Trqy \bigl(o(a_{[2k-h_a-n]}b )\bigr)\ , \quad n\geq 1\ , 
\ee
where only ordinary Eisenstein series $G_{2k}(q)$ appear. The 
square bracket modes $a_{[-l]}$ are defined in appendix~A. 

A new phenomenon appears if $a$ has $U(1)$ charge $Q\neq 0$, 
{\it i.e.}\ if $J_0 a = Q a \neq 0$: since the derivation
of the recursion relations involves cycling $a$ through the
trace, we pick up an additional factor $y^Q$ which 
enters the expression.
In particular, following the analysis
of \cite{Mason:2008zz} (see also appendix~A), one finds the recursion relation
\be
\Trqy \bigl(o(a_{[-h_a]}b) \bigr) 
= \sum_{k=1}^\infty \hat G_{k}(q,y^Q) 
    \Trqy \bigl(o(a_{[k-h_a]}b) \bigr)\ , \label{recQ1}
\ee
where $\hat G_k(q,y)$ are twisted Eisenstein series 
which we will define below. Note that this is only non-vanishing if $b$
has charge $-Q$.
Since we have the relation
$(L_{[-1]}a)_{[n]}= -(h_a+n-1)a_{[n-1]}$ we can obtain similar expressions
for lower modes, 
\be
{}\Trqy \bigl(o(a_{[-h_a-n]}b)\bigr) = (-1)^{n}\sum_{k\geq n+1}
\binom{k-1}{n} \hat G_k(q,y^Q)\, \Trqy \bigl(o(a_{[k-h_a-n]}b)\bigr) 
\ ,  \quad n\geq 1\ , 
\label{recQ2} 
\ee
which can be shown by  induction.
For future use we also note that by setting $b=\Omega$, 
we obtain $\Trqy \bigl(o(a_{[-n]}\Omega)\bigr)=0$ for $n > h_a$
for all operators $a$, charged or uncharged.

The twisted Eisenstein series $\hat G_k(q,y)$ that appear
in those expressions are defined for $k\geq1$ by 
\cite{Mason:2008zz} (see also \cite{Dong:1997ea})
\begin{equation}\label{genE}
\begin{array}{rcl}
{\displaystyle \hat G_{2k}(q,y)} &=& 
{\displaystyle 2\zeta(2k)+ \frac{(2\pi i)^{2k}}{(2k-1)!}
\sum_{n=1}^\infty \left[ \frac{n^{2k-1} q^n 
    y^{-1}}{1-q^ny^{-1}} + \frac{n^{2k-1} q^ny}{1-q^ny} \right] } \vspace{0.1cm} \\
\hat G_{2k+1}(q, y) &=& 
{\displaystyle \frac{(2\pi i)^{2k+1}}{(2k)!}
\sum_{n=1}^\infty \left[ \frac{n^{2k} q^n 
    y^{-1}}{1-q^ny^{-1}} - \frac{n^{2k} q^ny}{1-q^ny} \right] } \vspace{0.1cm} \\
\hat G_{1}(q, y) &=& 
{\displaystyle (2\pi i)
\sum_{n=1}^\infty \left[ \frac{ q^n 
    y^{-1}}{1-q^ny^{-1}} - \frac{q^ny}{1-q^ny} \right]
+\frac{2\pi i}{1-y^{-1}} -\pi i \ . }
\end{array}
\end{equation}
As usual, we will use the identification $q=e^{2\pi i \tau}$ and $y=e^{2\pi i z}$.
It is easy to see from the definitions that $\hat G_k(\tau, -z) =(-1)^k\hat G_k(\tau,z)$, so that
$\hat G_{2k+1}(\tau,0) = 0$. Furthermore,  we also recover
the usual Eisenstein series for $z=0$, 
$\hat G_{2k}(\tau,0) = G_{2k}(\tau)$.
Similar to the second Eisenstein series $G_2$, 
the generalised Eisenstein series $\hat G_k$ transform 
almost as forms of weight $k$,
up to anomalies. More precisely, they
are invariant under $\tau \mapsto \tau +1$, $z \mapsto z$, and under 
$\tau \mapsto -1/\tau$, $z \mapsto z/\tau$ they transform as
\be\label{transg}
(2\pi i)^{-m} \hat G_m \Bigl(-\frac{1}{\tau}, \frac{z}{\tau}\Bigr) = \sum_{k=1}^m
\frac{(-1)^{m-k}}{(m-k)!}(2\pi i)^{-k} \hat G_k(\tau,z) z^{m-k} \tau^k - (-1)^m
  \frac{z^m}{m!} \ . 
\ee
The transformation of $\hat G_1$ is particularly simple,
\be\label{trans1}
\hat G_1\Bigl(-\frac{1}{\tau},\frac{z}{\tau}\Bigr) = \tau \hat G_1(\tau,z) + (2\pi i)
z\ .
\ee
For the following it will be convenient to define rescaled Eisenstein series 
\be
{\G}_{2n} = \frac{1}{(2\pi i)^{2n}} G_{2n} \quad \hbox{and} \quad
\hat{\G}_{n} = \frac{1}{(2\pi i)^n} \hat{G}_n \ .
\ee

\section{Modular covariant differential equations}
\setcounter{equation}{0}

With these preparations we can now construct modular covariant
differential operators. Let us first review how this worked in the usual
bosonic case. It has been known for some time (for a precise statement
see for example \cite{Zhu}) that the torus one-point functions 
in the representation $\H_j$ of a state $\psi$
with $L_{[0]}$-eigenvalue $h$ 
\be\label{3.1}
\Tr_{\H_j} \Bigl( o(\psi) \, q^{L_0 - \frac{c}{24}} \Bigr)\ 
\ee
transform as a vector valued modular form of weight $h$. If we take $\psi$ to 
be of the form $u_l=L_{[-2]}^l\Omega$, then we can use the recursion relation
(\ref{rec01}) to rewrite (\ref{3.1}) in terms of $o(L_{[-2]}\Omega) \, o(u_{l-1})$, as well as 
one-point functions of states with lower conformal dimension. 
We can then replace the zero mode $o(L_{[-2]}\Omega)$ by the differential operator 
$D_q =q\frac{d}{dq}$ because
\be
o(L_{[-2]}\Omega)\, q^{L_0-\frac{c}{24}} = 
(2\pi i)^2 \left(L_0-\frac{c}{24}\right) q^{L_0-\frac{c}{24}} = 
(2\pi i)^2 \, D_q \, q^{L_0-c/24} \ .
\ee
By induction on $l$, we can then write the one-point function (\ref{3.1}) of
$\psi=u_l$ in terms of an $l$-th order differential operator acting on the 
character (see for example \cite{Gaberdiel:2008pr}). 
On the other hand, since we know that (\ref{3.1}) transforms as 
a modular form of weight $h$, the resulting differential operator must
be modular covariant of weight $h$. In fact, this can also be checked explicitly.
\medskip

For the case of the elliptic genus  
\be\label{3.3}
\Tr_{\R} \Bigl( o(\psi) \, (-1)^F\, y^{J_0}\, 
q^{L_0 - \frac{c}{24}} \Bigr)
\ee
the situation is more complicated because of the $y^{J_0}$
term. In fact, under the modular transformation
$\tau\mapsto -1/\tau$, $z\mapsto z/\tau$, the one-point function 
(\ref{3.3}) of a state $\psi$ with $L_{[0]}$ eigenvalue $h$ 
can be expressed in terms of the one-point function (\ref{3.3}) 
associated to the state $\exp(2\pi i z J_{[1]})\psi$ \cite{Eicher}. The
situation is therefore only simple provided that $\psi$ is annihilated 
by $J_{[1]}$, for which case (\ref{3.3}) transforms precisely as a weak
Jacobi form of weight $h$ and index $m=\frac{c}{6}$. (For the definition
of a weak Jacobi form see appendix~B.)

Using the recursion relations from section~2, we can convert the one-point
function of any Neveu-Schwarz 
$N=2$ descendant of the vacuum into a differential operator
(involving now derivatives with respect to $\tau$ and $z$) acting on the vacuum 
amplitude. Starting with a state that is annihilated by $J_{[1]}$ we can thus
construct modular covariant differential 
operators that act on weak Jacobi forms of weight zero. Unfortunately,
the analysis is significantly more complicated than in the bosonic case, 
and we cannot give a closed formula for all $h$. However, we can 
give explicit formulae for the first few cases, and we can count how many different
operators we may obtain in this manner.

\subsection{The operator of weight two}

At $h=1$ the only $N=2$ descendant of the vacuum is the state
$J_{[-1]}\Omega$, which is not annihilated by $J_{[1]}$. The first non-trivial
vector that is annihilated by $J_{[1]}$ appears at $h=2$. In fact, there are 
two such vectors: $J_{[-2]}\Omega$, which leads to a trivial differential equation, and 
\be \label{M2}
{\cal M}_2 = 
\left(L_{[-2]} - \frac{1}{2\tilde{c}}\, J_{[-1]} J_{[-1]} \right) \Omega \ ,
\ee
where $\tilde{c}=\tfrac{c}{3}$. Since ${\cal M}_2$ only involves the $L$ and $J$ modes 
that have $Q=0$, the old analysis of Zhu applies, and one finds after a short calculation 
that 
\be\label{3.5}
\Trqy (o({\cal M}_2))=  
(2\pi i)^2 \left( D_q - \frac{1}{4m} \, D_y^2 - \frac{1}{2} \,\G_2(q) \right) \, 
\Trqy ({\bf 1}) \ ,
\ee
where $D_y=y\frac{d}{dy}$ and $m=\tfrac{c}{6}=\tfrac{\tilde{c}}{2}$. We recognise the 
differential operator as the well known (modular covariant) 
heat kernel operator, whose action on a weak Jacobi form of 
weight $k$ and index $m$ is defined as 
\be\label{heat}
D_2^{(k)}= D_q - \frac{1}{4m} D^2_y + \frac{2k-1}{2} \, \G_2(q)  \ .
\ee
(In our case we have $k=0$, since the elliptic genus
on the right side of (\ref{3.5}) has weight zero.)

\subsection{The operator of weight three}

\noindent At $h=3$ there are (generically) three states that are annihilated by
$J_{[1]}$, namely 
\begin{eqnarray}
{\cal M}_{31} & = & J_{[-3]} \Omega \nonumber \\
{\cal M}_{32} & = & 
\left(\ct\, L_{[-3]}- J_{[-2]} J_{[-1]} \right)\Omega  \\
{\cal M}_{33} & = & 
\Bigl(- \frac{3}{2} \ct^2 \, G^-_{[- 3/2]} G^+_{[- 3/2]}
     - 3 \ct\, L_{[-2]} J_{[-1]}  
     + \frac{3}{2} \ct^2\, L_{[-3]}   + J_{[-1]}^3 \Bigr)\Omega \ .\nonumber 
\end{eqnarray}
Both ${\cal M}_{31}$ and ${\cal M}_{32}$
are $L_{[-1]}$ descendants, and hence give rise to trivial differential
operators since $o(L_{[-1]}b)=0$ for any state $b$. 
On the other hand, using the above recursion relations for 
${\cal M}_{33}$ (that now also involve charged modes), one finds
that the associated differential operator is 
\be
D_3= D_y^3+ 3\tilde c\, \G_2(q)\, D_y 
- 3\tilde c\, (D_q+ \G_2(q))\, D_y+ 3 \tilde c^2\Bigl( \hat \G_1(q,y)\, D_q 
+ \hat \G_2(q,y)\, D_y 
+ \tilde c \, \hat \G_3(q,y)\Bigr)\ .
\ee
Using the explicit modular 
transformation properties (\ref{transg}) and (\ref{trans1})
of the generalised Eisenstein series,
one can check that this differential operator really maps
a weak Jacobi form of weight zero and index $m=\tfrac{\tilde{c}}{2}$ to
one of weight three and index $m$.  
One can also show that $D_3$ annihilates the weak Jacobi form
$\Phi_{0,1}$, as must be the case since there are no weak Jacobi forms of 
weight/index $(3,1)$.

\subsection{The operators of weight four}

The analysis at $h=4$ is similar. The space of states that are annihilated by
$J_{[1]}$ is seven-dimensional, but the three vectors that are $L_{[-1]}$ descendants
of ${\cal M}_{31}$, ${\cal M}_{32}$ and ${\cal M}_{33}$ all lead to trivial 
differential operators. The same is true for the state 
\be
{\cal M}_{41} = \left( J_{[-2]} \, J_{[-1]}^2 - 2 \tilde{c}\, J_{[-2]} L_{[-2]} \right) \Omega \ .
\ee
Of the remaining states ${\cal M}_{42} = {J_{[-2]}}^2 \Omega$ 
simply leads to multiplication by $-6 \tilde{c}\, G_4(q)$. 
The state 
\be
{\cal M}_{43} = \left( L_{[-2]}^2 - \frac{1}{\tilde{c}}\, L_{[-2]} J_{[-1]}^2
+  L_{[-4]} + \frac{1}{4\tilde{c}^2}\, J_{[-1]}^4 \right) \Omega 
\ee
gives $D_2^{(2)} \cdot D_2^{(0)} + (\frac{3}{2}\, \tilde c-\frac{7}{2})\G_4(q)$,
which is the square of the heat kernel plus a $\G_4$ term.
Finally the state 
\begin{eqnarray}
{\cal M}_{44} & = & \Bigl[ (\tilde{c}-2)\, \left( G^+_{[-5/2]} G^-_{[-3/2]} 
+ G^-_{[-5/2]} G^+_{[-3/2]} \right)  + 2 G^-_{[-3/2]} G^+_{[-3/2]} J_{[-1]} \nonumber \\
& & \qquad  + 2 L_{[-2]}^2 - 2 L_{[-3]} J_{[-1]} + 2 L_{[-4]} \Bigr] \Omega  \ 
\end{eqnarray}
leads to the differential operator 
\begin{eqnarray}
D_4 & = &  2 \bigl(D_q + 2\G_2(q) \bigr)\, D_q + 3\tilde c \, \G_4(q) \\
& & + 4\hat\G_1(q,y)\left( \hat\G_1(q,y)D_q +\hat\G_2(q,y)D_y
+ \tilde c \hat\G_3(q,y) - (D_q+\G_2(q))D_y \right)  \nonumber \\
& & - 4\hat\G_2(q,y)\left( (\tilde c-2)D_q + D_y^2 +\tilde c \G_2(q)\right)
-12(\tilde c-1)\hat\G_3(q,y)D_y - 4\tilde c(3\tilde c -4)\hat \G_4(q,y) \ . \nonumber
\end{eqnarray}

\subsection{The general analysis}

Let us denote the Neveu-Schwarz $N=2$ vacuum representation by $\H$.
It is not difficult to determine the number of vectors in $\H$
 at fixed conformal weight and $U(1)$ charge that are generically annihilated by $J_{[1]}$; 
they are given by the generating function 
\begin{equation}
Z = 
\frac{\prod_{r=3/2,5/2,}^{\infty} \Bigl((1 + q^r\, y)  (1+ q^r y^{-1}) \Bigr) }
{\prod_{n=2}^{\infty} (1-q^n)^2 } \ .
\end{equation}
To obtain a non-vanishing operator, we need to take a state with
vanishing charge.
We are thus interested in the coefficients of $y^0$, which are given by 
\begin{equation}
Z[y^0] =  1 + 2 q^2 + 3 q^3 + 7 q^4 + 11 q^5 + 23 q^6 + 36 q^7 + 67 q^8 +
\cdots \ . 
\end{equation}
We have seen that not all states of a given conformal weight $h$ 
lead to non-trivial differential
operators. Some vectors, for example $J_{[-h]}\Omega$, will vanish identically, 
while others will lead to Eisenstein series multiplying differential operators of 
lower degree. However, it is easy to see that the highest derivatives that appear 
at that conformal weight are of the form 
\be\label{leading}
D_q^l \, D_y^{h-2l} \ , \qquad 
l=0,1,\ldots, \left\lfloor \frac{h}{2} \right\rfloor \ .
\ee
The vectors that give rise to these differential operators (plus correction terms
involving lower order derivatives) are the `leading vectors'
\be\label{leadv}
v_l^{(h)} = L_{[-2]}^l \, J_{[-1]}^{h-2l} \, \Omega \ , \qquad 
l=0,1,\ldots, \left\lfloor \frac{h}{2} \right\rfloor  \ .
\ee
They form a spanning set for the quotient space
$C_2^{\rm NS}=\H / O^{\rm NS}_{[2]}(\H)$, where $O_{[2]}^{\rm NS}(\H)$ is 
generated by the vectors\footnote{The space $O_{[2]}^{\rm NS}(\H)$ does not 
quite agree with the usual $O_{[2]}(\H)$ space, since it contains also the 
states generated by $G^\pm_{[-3/2]}$. Thus the resulting quotient space 
$C_2^{\rm NS}$ is also not quite the standard $C_2$ space. However, it 
gives a natural upper bound for the NS-sector version of Zhu's algebra.}
\be
O^{\rm NS}_{[2]}(\H)  = \hbox{span} \bigl\{L_{[-3-n]} \psi \ ,  J_{[-2-n]}\psi \ ,  G^\pm_{[-3/2-n]}\psi \ ,
n\geq 0 \bigr\} \ .
\ee
Now we want to show that for every conformal weight $h$ and modulo 
$O^{\rm NS}_{[2]}(\H)$, the space of vectors that are annihilated by $J_{[1]}$
has dimension $\lfloor \frac{h}{2} \rfloor$. 
This will imply that for every $h$, there are 
$\lfloor \tfrac{h}{2} \rfloor$ different modular covariant differential operators that 
differ by their leading terms (\ref{leading}). 

In fact, it is easy to see that the 
dimension of this space cannot be bigger than
$\lfloor \frac{h}{2} \rfloor$
since the dimension of $C_2^{\rm NS}$ at 
conformal weight $h$ is $\lfloor \frac{h}{2} \rfloor+1$, and since the condition
that $J_{[1]}$ annihilates the vector  implies that the term $J_{[-1]}^{h} \Omega$ 
always appears in the combination 
\be
{\cal M}_h = 
\left( L_{[-2]} J_{[-1]}^{h-2}  - \frac{1}{h  \tilde{c}} J_{[-1]}^h \right) \Omega + \hbox{other terms}\ . 
\ee
\smallskip

In order to prove that the space is at least as big as claimed,
we note from our previous explicit
calculations that at conformal weight $h=2,3,4$ there
are vectors that are annihilated by $J_{[1]}$, and whose leading terms are 
\begin{eqnarray}
{\cal M}_2 & = & \Bigl( L_{[-2]} - \frac{1}{2\tilde{c}} \, J_{[-1]}^2 \Bigr) \Omega \label{M2p} \\
{\cal M}_3 & = & \Bigl( L_{[-2]} J_{[-1]} - \frac{1}{3\tilde{c}} \, J_{[-1]}^3\Bigr)\Omega + 
O_{[2]}^{\rm NS}
\label{M3}  \\[4pt]
{\cal M}_4 & = & L_{[-2]}^2  \Omega + O_{[2]}^{\rm NS} \ .
\end{eqnarray}
One also finds that at $h=5$ there is a vector ${\cal M}_5$ that is annihilated by $J_{[1]}$, 
and that is of the form
\begin{equation}
{\cal M}_5 =  \Bigl( L_{[-2]} J_{[-1]}^3 - \frac{1}{5 \tilde{c}} J_{[-1]}^5 \Bigr) \Omega 
+ O_{[2]}^{\rm NS} \ .
\end{equation}
It is easy to see (by the same arguments as for the usual $C_2$ space) that the 
$C_2^{\rm NS}$ quotient space is a commutative algebra, where the product is
defined by $V_{[-h]}(\psi) \chi$, with $h$ the conformal weight of $\psi$.
Furthermore, if 
$\psi$ and $\chi$ are annihilated by $J_{[1]}$, so is their product $V_{[-h]}(\psi) \chi$.
Thus we can generate states that are annihilated by $J_{[1]}$ by taking successive products 
of  low-lying states. It only remains to check that the states we generate are sufficient in
number to prove the above claim. Using the commutative algebra structure on 
$C_2^{\rm NS}$ this is then effectively a problem in a polynomial algebra. Let us define the 
generators of $C_2^{\rm NS}$
\be
x = L_{[-2]} \Omega \ ,\qquad y = \sqrt{\frac{1}{2\tilde{c}}} J_{[-1]}\Omega \ ,
\ee
where $y$ has weight one, and $x$ has weight two. Then 
${\cal M}_2$, ${\cal M}_3$, ${\cal M}_4$ and ${\cal M}_5$ correspond to the 
generators 
\be
p_2 = (x - y^2)  \ , \qquad  p_3 = y \left(x-\frac{2}{3} y^2\right) \ , \qquad 
p_4 = x^2  \ , \qquad p_5 = y^3 \left(x-\frac{2}{5} y^2 \right) \ ,
\ee
respectively. It is easy to see that 
$p_2^3$, $p_2 p_4$ and $p_3^2$ generate at $h=6$ 
\be
p_6 = y^4 \left(x-\frac{1}{3} y^2\right) \ .
\ee
Let us first consider the case when $h$ is even, $h=2n$, and use
induction on $n$. For $h=2$ the statement is obviously true. Suppose
$f_1, \ldots, f_{n}$ are $n$ linearly independent polynomials of weight $2n$.
At weight $h=2n+2$ we then consider the polynomials
\be
\begin{array}{ll}
p_2 \cdot f_j, \quad  j=1,\ldots, n \ , \quad \hbox{and} \quad p_4^{\frac{n+1}{2}} & \qquad
\hbox{if $n$ is odd} \\
p_2 \cdot f_j,  \quad j=1,\ldots, n \ , \quad \hbox{and} \quad p_6 \cdot p_4^{\frac{n-2}{2}} & \qquad
\hbox{if $n$ is even.} 
\end{array}
\ee
Obviously, the polynomials $p_2 \cdot f_j$ are linearly independent in each case. By setting
$x=y^2$ they all vanish, but neither does $p_4$ nor $p_6$. Thus the last element in each
case is linearly independent of the first $n$ elements, and we have therefore found 
$\frac{h}{2}=n+1$ linearly independent polynomials.

For odd $h$ the argument is essentially the same. In constructing the polynomials 
at $h=2n+1$ from those at $2n-1$ we add either $p_3 \cdot p_4^m$ with $m=\tfrac{n-1}{2}$
(if $n$ is odd) or $p_5 \cdot p_4^{m}$ with $m=\tfrac{n-2}{2}$ (if $n$ is even). This completes
the proof.

\section{Applications}
\setcounter{equation}{0}

Finally, let us discuss some applications of these general considerations.

\subsection{Modular differential equations of minimal models}

For the $N=2$ minimal models one expects that differential operators
of the kind discussed above should be the building blocks for the 
modular differential equation that characterises all elliptic genera. Recall
that the central charge of the $k^{\rm th}$ minimal model is given by 
\be
c = \frac{3k}{k+2}\ .
\ee
The elliptic genera of these models were  determined in \cite{Witten:1993jg,Kawai:1993jk}.
Let us first consider the case $k=1$ for which $c=1$, so that $\tilde{c}=\tfrac{1}{3}$ 
and $m=\tfrac{1}{6}$. In this case, the vector
${\cal N}= {\cal M}_2$ given in (\ref{M2}) is actually a null vector. This
means that the corresponding differential operator
must annihilate the elliptic genera,
\be
\left( D_q -\frac{3}{2}D_y^2 -\frac{1}{2} \G_2(q) \right) \chi_{\cal R}(q,y) 
= 0\ .\label{k1eq1}
\ee
A second null vector of the theory can be obtained by applying 
$G^-_{[-1/2]} G^+_{[-1/2]}$ to ${\cal N}$, leading to 
\be
\hat{\mathcal{N}} = \bigl(3G^-_{[-3/2]}G^+_{[-3/2]} - 2L_{[-3]} + 6L_{[-2]}J_{[-1]}
-3J_{[-2]}J_{[-1]} -4J_{[-3]}\bigr)\Omega\ .
\ee
Note that this is a linear combination of ${\cal M}_{31}$, ${\cal M}_{32}$ and 
${\cal M}_{33}$, as well as $J_{[-1]} {\cal N}$. It leads to the differential equation
\be
\Bigl( (D_q + \G_2(q))D_y - \hat \G_1(q,y) D_q
 - \hat \G_2(q,y) D_y - \frac{1}{3} \hat \G_3(q,y)\Bigr ) \,
 \chi_{\cal R}(q,y) = 0\  . \label{k1eq2}
\ee
The $k=1$ minimal model has three elliptic genera,
\ba\label{c1a}
\chi^0_1(q,y) &=& y^{-1/6} + (y^{-1/6} - y^{5/6})q + 
(-y^{-7/6} +2y^{-1/6} -y^{5/6})q^2 + \cdots \nonumber \\
\chi^0_{-1}(q,y) &=& -y^{1/6} + (-y^{1/6} + y^{-5/6})q + 
(y^{7/6} -2y^{1/6} +y^{-5/6})q^2 + \cdots \\
\chi^0_3(q,y) &=& (-y^{-1/2} + y^{1/2})q^{1/3} + (-y^{-1/2} + y^{1/2})q^{4/3}
+\cdots\ . \nonumber
\ea
It is straightforward to check that these are indeed annihilated
by the two differential equations. 

Conversely, one can also show that (\ref{c1a}) 
are the only solutions of (\ref{k1eq1}) and (\ref{k1eq2}). To this end we 
make a power series ansatz for $\chi(q,y)$,
\be\label{ansatz}
\chi(q,y) = q^h y^Q \sum_{\ell=0}^N c(\ell) y^\ell  + {\cal O}(q^{h+1}) 
\ , \qquad c(0)\neq 0\ .
\ee
Since the differential operators never decrease the power of $q$, 
the terms we have written out explicitly in (\ref{ansatz}) 
must be annihilated by the differential operators up to terms of order 
${\cal O}(q^{h+1})$. Acting with (\ref{k1eq1}) we thus obtain 
\be
h = \frac{3}{2}\, Q^2-\frac{1}{24} \ , \label{constraint1}
\ee
while (\ref{k1eq2}) leads to
\be \label{constraint2}
h (1+2Q) = 0\ .
\ee
Note that there is a slight subtlety in deriving the second equation because
$\hat\G_1$ contains the term
$(1-y)^{-1}$. However, since the elliptic genus is an index we know that 
$\chi(q,y)$  must vanish for $y=1$ unless $h=0$. 
This means that either $D_q$ vanishes or that we can factor out $(1-y)$,
so that the action of $\hat \G_1$ is indeed well-defined.

The two equations (\ref{constraint1}) and (\ref{constraint2}) must be 
satisfied simultaneously, and their only solutions are 
$(h=0,Q=\pm\frac{1}{6})$ and $(h=\tfrac{1}{3},Q=-\tfrac{1}{2})$.
This shows that there are at most three solutions of the form (\ref{ansatz}), and thus,
given the existence of the explicit solutions  $\chi_{\pm 1}^0$ and $\chi_3^0$, 
precisely three such solutions.
\medskip

The analysis for a general minimal $N=2$ model is similar. The 
$k^{\rm th}$ minimal model has an uncharged null vector $\N_k$ at level $k+1$,
and the non-trivial descendant $G^-_{[-1/2]}G^+_{[-1/2]} \N_k$ at level $k+2$. 
Acting on the lowest powers of $q$ these will again give two polynomial
equations for $h$ and $Q$ whose leading terms have weight $k+1$
and $k+2$ respectively. (Here we associate weight one to $Q$, and weight two to
$h$.) To find the number of joint solutions, first substitute
$h=H^2$, so that $H$ and $Q$ have both weight one. In addition, we introduce
an auxiliary variable $z$ of weight one, and multiply the different terms so as 
to obtain homogeneous
polynomials of degree $k+1$ and $k+2$, respectively. 
These equations can now be thought of as equations in projective space
$\mathbb{P}^2$, and B\'ezout's theorem then implies that there are $(k+1)(k+2)$ 
solutions, taking into account multiplicities. Since they obviously come in pairs of $(Q,\pm H)$, 
it follows that there are at most $(k+1)(k+2)/2$ solutions for $(Q,h)$. This is exactly the 
number of independent elliptic genera of the $k^{\rm th}$ minimal model.

\subsection{Modular constraints for extremal self-dual theories}

Modular differential equations gave an interesting constraint for extremal (bosonic)
conformal field theories \cite{Gaberdiel:2007ve,Gaberdiel:2008pr}. Given our 
insight into the structure of differential operators for weak Jacobi forms, we 
can now apply similar techniques to the $N=2$ case. To start with we recall that 
the space of weak Jacobi forms of even weight $w$ and index $m$,
$\tilde{J}_{m,w}$,  is spanned by monomials of the form
(see appendix~B)
\be
(\tilde \phi_{-2,1})^a\, (\tilde \phi_{0,1})^b \, G_4^c \, G_6^d \ ,
\ee
where $a+b=m$ and $4c+6d-2a = w$. The number of solutions
to the second equation is the number of modular forms of weight $w'=w+2a$, which
is proportional to 
\be
N(w') = \frac{w'}{12} + \textrm{subleading} \ .
\ee
Thus the total dimension is, for large $m$, of the form 
\be
\dim \tilde J_{m,w} = \frac{m^2}{12} + \frac{wm}{12} + \textrm{subleading} \ .
\ee
Suppose now that we are given a self-dual $N=2$ superconformal field theory 
at $c=6m$. Its elliptic genus defines a weak Jacobi form of weight zero and index 
$m$. We want to construct a modular differential equation of weight $w$ that annihilates 
it. As we have seen, there are $\lfloor\frac{w'}{2}\rfloor$ 
differential operators of order $w'$. Dressing
them with an ordinary modular form of weight $w-w'$, the total number of
differential operators of degree that we can construct in this manner is, for large $w$,
\be
\frac{1}{2}\sum_{w'=0}^{w}  \frac{w'}{2} \frac{w-w'}{12} = \frac{w^3}{288} + \textrm{subleading}\ ,
\ee
where the factor $\frac{1}{2}$ corrects for the fact that there are no modular
forms of odd weight.
If we apply any such differential operator to 
the elliptic genus of our given $N=2$ superconformal field theory
at $c=6m$, we obtain a weak Jacobi form of weight $w$ and index $m$.
Since there are at most $\dim(\tilde{J}_{m,w})$ linearly independent weak Jacobi
forms of this weight and index, we can always choose a suitable linear combination 
of the differential operators to annihilate the elliptic genus if 
\be
\frac{w^3}{288}  \geq \frac{m^2}{12} + \frac{wm}{12} \ ,
\ee
\ie\ if $w\geq w_{*}$ with
\be\label{N2a}
w_{*} = 24^{\frac{1}{3}} \,  m^{\frac{2}{3}} = \bigl(\tfrac{2}{3}\bigr)^{\frac{1}{3}} \, c^{\frac{2}{3}} \ .
\ee
Following the logic of the arguments of \cite{Gaberdiel:2007ve,Gaberdiel:2008pr},
this would then suggest that the self-dual $N=2$ superconformal field theory has a null
vector at conformal weight $w_*$. Since the superconformal descendants of the 
vacuum do not have any null vectors
for $c>3$, this would thus imply that the theory
has to have additional primary fields at conformal weight $h\sim w_{*}$. 

Note that we could also ignore the $N=2$ supersymmetry of the problem,
and treat the theory from a bosonic point of view. With respect
to the bosonic generators, the theory is not self-dual, and 
the original counting argument needs to be modified slightly; however, this
will still lead to a differential equation of weight $w_* \sim \sqrt{2c}$, which 
is stronger than (\ref{N2a}). On the other hand, one may hope that 
the more refined $N=2$ analysis from above will help us to fill the 
gap in the argument of \cite{Gaberdiel:2008pr}, and thus prove a 
no-go theorem in this case.

\section*{Acknowledgements}
We thank Sergei Gukov for collaboration at an early stage of this
project. We are also very grateful to Claude Eicher, Terry Gannon,
Sebastian Gerigk, Roberto Volpato and Don Zagier
for helpful discussions. The research of M.R.G.\  
is supported by the Swiss National Science Foundation, and 
C.A.K.\ is supported by a Fellowship of 
the Swiss National Science Foundation.

\appendix
 \renewcommand{\theequation}{\Alph{section}.\arabic{equation}}

\section{Conventions and recursion relations}\label{app:zhuN2}
\setcounter{equation}{0}

\subsection{Conventions}

Let us begin by collecting our conventions.
The vertex operator of a state $a$ is given by
\be\label{modex}
V(a,w) = \sum_n a_n \, w^{-n-h_a} \ .
\ee
Note that we use a different convention than \cite{Zhu} --- in
particular, $a_n = a(h+n-1)$. It is sometimes convenient to 
introduce a special symbol for the zero mode as 
\be
o(a) = a_0 \ . 
\ee
The modes that are appropriate for the analysis on the torus are defined by
(see section 4.2 of \cite{Zhu}) 
\be
V[a,z] = e^{2\pi i z h_a}\, V(a,e^{2\pi i z}-1) = \sum_n a_{[n]}
z^{-n-h_a} \ . \label{trafo} 
\ee
The different modes are related to one another via
\be \label{bracketmodes}
a_{[m]} = (2\pi i)^{-m-h_a}\sum_{j\geq m} c(h_a,j+h-1,m+h-1) a_j \ , 
\ee 
where 
\be
(\log(1+z))^m (1+z)^{h_a-1} = \sum_{j\geq m} c(h_a,j,m) z^j \ . 
\ee
This defines a new vertex operator algebra with an energy tensor whose
modes $L_{[n]}$ are given by
\be
L_{[n]}= (2\pi i)^{-n}\sum_{j\geq n+1} c(2,j,n+1) L_{j-1} -
(2\pi i)^2\frac{c}{24}\delta_{n,-2} \ .
\ee
The second term on the right hand side comes from the fact that $L$ is only
quasiprimary and thus picks up an anomaly under the coordinate
transformation (\ref{trafo}). Note that compared to (\ref{bracketmodes})
we have chosen a slightly
different overall normalisation so as to obtain
the standard commutation relations for the modes $L_{[n]}$.
To maintain the standard $N=2$ commutation relations 
we also choose an analogous normalisation for the $J_{[n]}$ and $G^\pm_{[n]}$, 
\ie without the prefactor $(2\pi i)^{-h_a}$. With these conventions
their commutation relations are
\begin{align*}
  \left[L_{[m]},L_{[n]} \right] &=(m-n)L_{[m+n]} + \tfrac{\tilde{c}}{4} m\, (m^2 -  1) \,
      \delta_{m,-n} \ , \\ 
  \left[L_{[m]}, J_{[n]} \right] &=-nJ_{[m+n]}\ ,\\
  \left[L_{[m]}, G_{[n]}^\pm \right] &=\left( \tfrac{1}{2}m-n \right)
  G_{[m+n]}^\pm \ , \\
  \left[J_{[m]}, J_{[n]} \right] &= \tilde{c}\, m\, \delta_{m,-n}\ ,\\
  \left[J_{[m]}, G_{[n]}^\pm \right] &=\pm G_{[m+n]}^\pm \ , \\
  \left\{ G_{[m]}^+, G_{[n]}^-\right\} &= 2 L_{[m+n]} + (m-n) J_{[m+n]} +
  \tilde{c}\, (m^2 - \tfrac{1}{4})\delta_{m, -n} \ , \\
  \left\{G_{[m]}^+, G_{[n]}^+ \right\} &=
   \left\{G_{[m]}^-, G_{[n]}^-\right\}=0\ ,
\end{align*}
as well as 
\be
o(J_{[-1]}\Omega) = (2\pi i) J_0 \ .
\ee

\subsection{Recursion relations}

Zhu's original argument \cite{Zhu} was generalised to superalgebras
in \cite{Mason:2008zz}. We will adapt it in the following 
to the $N=2$ case to extract the recursion relation (\ref{recQ1}) 
we shall need.

We define the elliptic genus amplitude by 
\be\label{egdef}
G_\R\Bigl( (a^1,z_1),\ldots,(a^m,z_m);q,y \Bigr) = 
\prod_{i=1}^{m} z_i^{h_i} \, 
{\rm Tr}_\R \Bigl( V(a^1,z_1) \cdots V(a^m,z_m) \,
q^{L_0} \, y^{J_0}\, (-1)^F \Bigr) \ ,
\ee
where $h_i$ is the conformal weight of the state $a^i$ (with respect to 
$L_0$), and $\R$ is the (irreducible) Ramond sector representation 
in which the trace is being taken. 

There are two key identities from which the recursion relations 
can be deduced. The first identity is 
\begin{eqnarray}
& & G_\R\Bigl( (a,w), (a^1,z_1),(a^2,z_2),\ldots,(a^n,z_n);q,y \Bigr)
\nonumber \\
& & \qquad \quad = 
\sum_{j=1}^{n}\,\Bigr[ \pm \sum_{m\in\Nop_0} \, 
\hat {\cal P}_{m+1}\left(\frac{z_j}{w},q,y^Q\right) \label{prop1} \\
& & \qquad \qquad \qquad \qquad \times 
G_\R\Bigl( (a^1,z_1),(a^2,z_2),\ldots,
(a_{[m-h_a+1]}a^j,z_j),\ldots , (a^n,z_n);q,y \Bigr)\Bigr] \ .
\nonumber
\end{eqnarray}
Here $a$ is a state with non-vanishing $U(1)$-charge, $J_0 a = Q a \neq 0$, 
and the function $\hat{\cal P}_k(q_w,q,y)$ is defined in appendix
\ref{app:Eisenstein}. It is meromorphic in $q_w$ and has a pole at
$q_w=1$. The $\pm$-signs in (\ref{prop1}) depend in the obvious manner
on whether $a$ and the other fields $a^j$ are fermionic or
bosonic; in particular, if $a$ is bosonic, there are no minus signs.  
It should also be clear that (up to the obvious $\pm$ signs) the elliptic
genus is independent of the order in which the fields appear.

The proof of (\ref{prop1}) is a straightforward extension of the original
argument due to Zhu \cite{Zhu} (see also the proof of
Theorem 3.6 in \cite{Mason:2008zz}). The basic
idea is to expand the vertex operator $(a,w)$ in terms of modes  as in 
(\ref{modex}).  Each mode is then commuted (or anti-commuted) through the 
trace until it returns to its place; since it picks up a non-trivial factor upon
going through $y^{J_0}$ and $q^{L_0}$, we can solve for it in terms of the commutators and 
anti-commutators; this leads to the recursion relation (\ref{prop1}).
\medskip

The second key identity is 
\begin{eqnarray}
& & G_\R\Bigl( (a_{[-h_a]}a^1,z_1),(a^2,z_2),\ldots,(a^n,z_n);q,y \Bigr)
\nonumber \\
& & \qquad \qquad = \sum_{k=1}^{\infty} \hat G_{k}(q,y^Q) \, 
G_\R\Bigl( (a_{[k-h_a]}a^1,z_1),(a^2,z_2),\ldots,(a^n,z_n);q,y \Bigr)
\nonumber \\
& & \qquad \qquad \quad + \sum_{j=2}^{n} \pm \sum_{m\in\Nop_0}
\hat {\cal P}_{m+1}\left(\frac{z_j}{z_1},q,y^Q\right)  \label{prop2} \\
& & \qquad \qquad \qquad \qquad \times
G_\R\Bigl( (a^1,z_1),(a^2,z_2),\ldots,(a_{[m-h_a+1]}a^j,z_j),\ldots ,
(a^n,z_n);q,y \Bigr) \ ,
\nonumber
\end{eqnarray}
where $\hat G_{k}(q,y)$ is the generalised Eisenstein series defined in
appendix \ref{app:Eisenstein}.

Since the proof in \cite{Mason:2008zz} is rather sketchy, we will spell it out
in more detail. First note that $a_{[-h_a]} a^1$
is not homogeneous with respect to $L_0$. In order
to insert it nevertheless into the elliptic genus we extend the definition of 
(\ref{egdef}) by linearity. We expand the first argument of $G_\R$ in (\ref{prop2})
to get
\begin{eqnarray}
(a_{[-h_a]} a^1,z_1) & = & \sum_{i\geq -1} c_i\, (a_{i-h_a+1} a^1,z_1) 
= \sum_{i\geq -1} c_i \, {\rm Res}_{w-z_1} 
(w-z_1)^i V\left( V(a,w-z_1) a^1,z_1 \right)  \nonumber \\
& = & \sum_{i\geq -1} c_i \, {\rm Res}_{w-z_1}
(w-z_1)^i V(a,w)\, V(a^1,z_1)  \ ,\nonumber 
\end{eqnarray}
where we have defined $c_i=c(h_a,i,-1)$.   
Taking into account the additional factors of $z_1$ and $w$ that come from 
the definition (\ref{egdef}) of $G_\R$ we obtain the sum
\be
\sum_{i\geq-1}c_i \,(w-z_1)^i z_1^{h_a-1-i}w^{-h_a}=
w^{-1}\left(\log\left(\frac{w}{z_1}\right)\right)^{-1}\  .
\ee
We now apply (\ref{prop1})
to our expression, obtaining terms of the form 
\begin{multline}
G_\R\Bigl( (a^1,z_1),\ldots,(a_{[m-h_a+1]}a^j,z_j),\ldots ,
(a^n,z_n);q,y \Bigr)\\
\times
\int_C 
w^{-1}\left(\log\left(\frac{w}{z_1}\right)\right)^{-1} \hat {\cal
  P}_{m+1}\left(\frac{z_j}{w},q,y^Q\right) dw\ ,  
\end{multline}
where the contour $C$ is around $z_1$.

For $j\neq 1$, $\hat{\cal P}_{m+1}\left(\frac{z_j}{w},q,y^Q\right)$
is regular at $w=z_1$, so that we obtain directly the last two lines
of (\ref{prop2}). For $j=1$, however, $\hat {\cal P}_{m+1}(\frac{z_1}{w},q,y^Q)$ 
has a pole at $w=z_1$. 
We perform the change of variable $w = z_1 \exp(-2\pi i w')$, which yields
$\textrm{Res}_{w'} \hat{\cal P}_{m+1}(e^{2\pi i w'},q,y^Q)$. Using
expansion (\ref{PQexpansion}) then gives the first term of (\ref{prop2}). 
\smallskip

The recursion relation (\ref{recQ1}) then follows directly from
(\ref{prop2}) by comparing the $z^0$ coefficient for $n=1$.

\section{Weak Jacobi forms}\label{app:weakJacobi}
\setcounter{equation}{0}
A weak Jacobi form \cite{Eichler} of weight $k$ and index $m$ is a function 
$f(\tau,z)$ on $\mathbb{H}_+\times \mathbb{C}$ 
satisfying the transformation property 
\be
f\left(\frac{a\tau + b}{c\tau + d} , \frac{z}{c\tau + d} \right) = 
(c\tau + d)^k\, e^{\frac{2\pi i m c z^2}{c\tau + d}} \, f(\tau,z) \ , 
\qquad \left(
\begin{array}{cc}
a & b \cr c & d 
\end{array}
\right) \in {\rm SL}(2,\Zop) \ ,
\ee
and
\be
f(\tau,z+r \tau + s) = e^{-2\pi i m (r^2\tau + 2 r z) }
f(\tau,z) \ , 
\ee
with $r,s\in\Zop$. Moreover it must have a Fourier expansion of the form 
\be
f(\tau,z) = \sum_{n=0}^\infty \sum_{l\in\Zop} c(n,l) \, q^n\, y^l \ , 
\ee
where as usual
\be
q = e^{2\pi i \tau} \ , \qquad y = e^{2\pi i z} \ .
\ee
The space of all weak Jacobi forms is generated by the two Eisenstein
series $G_4(q)$ and $G_6(q)$ (which are conventional modular forms of
weight $4$ and $6$, respectively), and the two weak Jacobi forms
\be
\phi_{-2,1} = \frac{\phi_{10,1}}{\Delta} \ , \qquad
\phi_{0,1} = \frac{\phi_{12,1}}{\Delta} \ , 
\ee
where $\Delta = q\prod_{n=1}^{\infty} (1-q^n)^{24}$ and
\be
\phi_{10,1} = \frac{1}{576\,\zeta(4)\zeta(6)} (G_6 G_{4,1} - G_4 G_{6,1}) \ , \qquad
\phi_{12,1} = \frac{1}{576} \left(\frac{G_4^2 G_{4,1}}{2\zeta(4)^3} 
- \frac{G_6 G_{6,1}}{\zeta(6)^2}\right) \ .
\ee
Here $\zeta(2k)$ is the Riemann zeta function and 
$G_{4,1}$, $G_{6,1}$ are Jacobi forms defined in \cite{Eichler}. 
On general grounds, one can show that the elliptic genus of an $N=2$
conformal field theory of central charge $c$ is a weak Jacobi form of
weight $0$ and index $m=\frac{c}{6}$.

\section{Twisted Eisenstein series}
\setcounter{equation}{0} \label{app:Eisenstein}
\subsection{Ordinary Eisenstein series}
The ordinary Eisenstein series are defined by
\ba
G_{2k}(\tau)&=& \sum_{(m,n)\neq(0,0)} \frac{1}{(m\tau +n)^{2k}} \qquad
  k\geq 2\ ,\\
G_{2}(\tau) &=& \frac{\pi^2}{3} + \sum_{m\in\mathbb{Z}-\{0 \}}
\sum_{n\in\mathbb{Z}} \frac{1}{(m\tau+n)^2}\ . 
\ea
They can also be written as
\be
G_{2k}(\tau) = 2\zeta(2k)+ \frac{2(2\pi i)^{2k}}{(2k-1)!}
\sum_{n=1}^\infty \frac{n^{2k-1} q^n}{1- q^n}\ .
\ee
For $k\geq 2$, $G_{2k}(\tau)$ is a modular form of weight $2k$, \ie
\be
G_{2k}\left(\frac{a\tau+b}{c\tau+d}\right) 
= (c\tau +d)^{2k}G_{2k}(\tau)\ ,
\ee
whereas $G_2(\tau)$ transforms as
\be
G_{2}\left(\frac{a\tau+b}{c\tau+d}\right) 
= (c\tau +d)^{2}G_{2}(\tau) - 2\pi i c(c\tau+d) \ .
\ee

\subsection{Twisted Eisenstein series}
For $|q|<|q_w|<1$ and $y\neq 1$ define
\be
\hat {\cal P}_m(q_w,q,y) := \frac{(2\pi i)^m}{(m-1)!} \, 
\left( \sum_{n=1}^{\infty} \frac{n^{m-1} q_w^n}{1 - q^n \, y^{-1}} 
+ \frac{(-1)^m n^{m-1} q_w^{-n} q^n \, y}
{1 - q^n \, y} \right) + \delta_{1,m}\, \frac{2\pi i}{1-y^{-1}}  \ .
\ee
Note that $\hat {\cal P}_m$ is a special case of the twisted Weierstrass
function $P_m$ introduced in \cite{Dong:1997ea,Mason:2008zz}. More precisely,
\be
\hat {\cal P}_m(q_w,q,y) = (-2\pi i)^m P_m \left[
\begin{array}{c} y^{-1} \\ 1 \end{array}\right](w,\tau) \ .
\ee

$\hat {\cal P}_m(q_w,q,y)$ converges for $|q|<|q_w|<1$, and
\be \label{delPm}
\frac{\partial}{\partial w}\hat{\cal P}_m(w,q,y) = 2\pi i
q_w \frac{d}{dq_w} \hat{\cal P}_m(w,q,y) = m\hat {\cal P}_{m+1}(w,q,y)\ .
\ee 
\noindent 
$\hat{\cal P}_1(q_w,q,y)$ has a simple pole at
$q_w=1$, as 
can be seen by rewriting
\be
\hat{\cal P}_1(q_w,q,y) = \frac{2\pi i}{1-q_w} - 2\pi i + \frac{2\pi
  i}{1-y^{-1}} 
+ 2\pi i
\sum_{n=1}^\infty \left( \frac{q_w^n q^n y^{-1}}{1-q^ny^{-1}} -
  \frac{q_w^{-n}q^n y}{1-q^ny}\right) \ , \label{Phat}
\ee
where the sum is seen to be convergent for $|q|<|q_w|<|q|^{-1}$. 
Choose $q_w$ so that $1<|q_w|<|q|^{-1}$. The series then converges for
both $q_w$ and $q q_w$, and a straightforward calculation shows the
identity
\be
\hat{\cal P}_1(qq_w,q,y) = y\,\hat{\cal P}_1(q_w,q,y)\ . \label{Qperiod}
\ee
Writing $q_w\equiv e^{2\pi i w}$, we expand (\ref{Phat}) in $w$. The
last terms give
\be
2\pi i \sum_{n=1}^\infty \frac{1}{k!}(2\pi i)^k \left( \frac{n^k q^n
    y^{-1}}{1-q^n y^{-1}}- \frac{(-n)^k q^n y}{1-q^ny}\right)
\times\, w^k\ ,
\ee
whereas the first term is
\be
-\frac{1}{w}\sum_{n=0}^\infty \frac{B_n}{n!} (2\pi i w)^n =
-\frac{1}{w} + \pi i + \sum_{n=1}^\infty 2\zeta(2n)\times w^{2n-1} \ .
\ee
Here we have used
\be
\frac{x}{e^x-1} = \sum_{n=0}^\infty \frac{B_n}{n!}x^n\ ,
\ee
where the Bernoulli numbers $B_n$ are given by $B_0=1$, 
$B_1 = -\frac{1}{2}$, and for $n\geq 1$, $B_{2n+1}=0$ and 
$B_{2n}=(-1)^{n-1}2(2n)!(2\pi)^{-2n} \zeta(2n)$.
In total, we thus obtain 
\be
\hat {\cal P}_1(w,q,y) =  -\frac{1}{w} + \hat G_1(q,y) +
\sum_{k=2}^\infty \hat G_k(q,y) w^{k-1}\ , \label{PQexpansion}
\ee
where we have introduced the twisted Eisenstein series
$\hat G_k(q,y)$ defined in (\ref{genE}).
We can then use identity (\ref{delPm})
to extract the poles of $\hat {\cal P}_m$.

\subsection{Transformation properties of the twisted Eisenstein series} 

In order to determine the modular behaviour of $\hat G_k$, it is sufficient to
consider the actions $\tau \mapsto \tau +1\ , z \mapsto z$ and $\tau
\mapsto -1/\tau\ , \ z \mapsto z/\tau$. $\hat G_k$ is obviously invariant
under the first, while under the second we claim that the transformation is 
\be
(2\pi i)^{-m} \hat G_m(-\frac{1}{\tau}, \frac{z}{\tau}) = \sum_{k=1}^m
\frac{(-1)^{m-k}}{(m-k)!}(2\pi i)^{-k} \hat G_k(\tau,z) z^{m-k} \tau^k - (-1)^m
  \frac{z^m}{m!} \ . \label{G1mtrafo}
\ee
Note that this is very similar to the
transformation properties of $G_2(\tau)$: it transforms almost as a form
of weight $m$ and index 0, but has additional anomalous terms.
To prove (\ref{G1mtrafo}), let us assume for the moment that $\hat G_1$
transforms as 
\be
\hat G_1(-\frac{1}{\tau},\frac{z}{\tau}) = \tau \hat G_1(\tau,z) + (2\pi i)
z\ . \label{G11trafo}
\ee

We can then prove (\ref{G1mtrafo}) by recurrence
using
\be
\frac{\partial}{\partial z} \hat G_{m+1}(\tau, z) = -\frac{1}{m} 
\frac{\partial}{\partial \tau} \hat G_{m}(\tau, z)\ . 
\ee
Introducing variables $\tilde \tau = -1/\tau\ , \ \tilde z = z/\tau$ and 
using 
\be
\partial_{\tilde \tau} = \tau^2 \partial_\tau + \tau z \partial_z\ , 
\qquad \partial_{\tilde z} = \tau \partial_z\ ,
\ee
we obtain
\be
\partial_z \hat G_{m+1}^1(-\frac{1}{\tau}, \frac{z}{\tau})
 = -\frac{1}{m} 
(\tau \partial_\tau +z \partial_z) \hat G_{m}(-\frac{1}{\tau}, \frac{z}{\tau})\ .
\ee
A straightforward calculation shows that this is equal to $\partial_z$
of the right hand side of (\ref{G1mtrafo}) for $m+1$. We have thus
shown that (\ref{G1mtrafo}) holds up to some function $f(\tau)$. To see
that $f=0$, consider the limit $z \rightarrow 0$. Since
$\hat G_{2k}(\tau,0)=G_{2k}(\tau)$, the known transformation properties
of the ordinary Eisenstein series fix $f=0$ for the even case. In
the odd case the same is true since $\hat G_{2k+1}(\tau,0)=0$. The only
somewhat subtle case occurs for $\hat G_2$. In this case
$\lim_{z\rightarrow 0} z \hat G_1(\tau,z)= 1$ produces the desired
modular anomaly so that again $f=0$.

Let us briefly sketch how to prove (\ref{G11trafo}). One can show that
$\hat G_1$ can be rewritten as
\be
\hat G_1(\tau, z) = - \sum_{n > 0} \sum_{k>0} S_{nk}
- \sum_{n>0}\frac{2z}{(n\tau)^2-z^2} - \sum_{k>0}\frac{2z}{k^2-z^2} + 
\frac{1}{z}\  ,\label{G11sum}
\ee
where
\be
S_{nk}= 
 \frac{2z}{(k+n\tau)^2 - z^2} + \frac{2z}{(-k+n\tau)^2 - z^2}\ .
\ee
Taking $\tau \mapsto -1/\tau$, $z \mapsto z/\tau$, the last three terms 
of (\ref{G11sum}) transforms to $\tau$ times themselves. The first sum
transform as 
\be
\sum_{n>0} \sum_{k>0} S_{nk} \mapsto \tau \sum_{n>0} \sum_{k>0} S_{kn}
= \tau \sum_{k>0} \sum_{n>0} S_{nk}
\ee
which is formally $\tau$ times the original expression. Since the
series is not absolutely convergent however, the order of summation
matters. By using standard methods one can show that exchanging the
summation order produces exactly the modular anomaly of 
(\ref{G11trafo}).

Another possibility is to derive (\ref{G1mtrafo}) directly from 
the transformation properties of the twisted Weierstrass functions
\cite{Mason:2008zz}.


\end{document}